\documentclass[11pt]{article}
\usepackage{moriond,epsfig,pennames}

\def\Journal#1#2#3#4{{#1} {\bf #2}, #3 (#4)}

\def\PLB{{\em Phys. Lett.}  B}

\def\PRD{{\em Phys. Rev.} D}

\def\JHEP{{\em J. High Energy Phys.}}
\def\NJP{{\em New J. Phys.}}
\def\JINST{{\em JINST}}

\def\be{\begin{equation}}
\def\ee{\end{equation}}
\def\bea{\begin{eqnarray}}
\def\eea{\end{eqnarray}}

\begin{document}
\vspace*{4cm}
\title{SOFT QCD RESULTS FROM ATLAS AND CMS
}

\author{ CLAUDIA-ELISABETH WULZ, for the ATLAS and CMS Collaborations }

\address{Institute of High Energy Physics of the Austrian Academy of Sciences,\\
Nikolsdorfergasse 18, 1050 Vienna, Austria}

\maketitle\abstracts{
The ATLAS and CMS Collaborations have measured properties of minimum bias events and have determined characteristics of the underlying event in proton-proton collisions at three LHC centre-of-mass energies. 
Comparisons to common phenomenological models and partially to other experiments have been made. The production of the strange particles $\PKzS, \Lambda$ and $\Xi$ is discussed. Particle correlation studies, in particular Bose-Einstein as well as long- and short-range angular correlations in proton-proton and lead ion events are explained.
}

\section{Properties of minimum bias events}
Ideally minimum bias events are those recorded with a totally inclusive trigger. The exact definition depends on the experiment.
Usually minimum bias only refers to non-single-diffractive (NSD) events. In ATLAS \cite{ATLASdet} and CMS \cite{CMSdet} similar minimum bias trigger detectors are used. ATLAS has two stations of Minimum Bias Trigger Scintillators (MBTS) located upstream and downstream at $z = \pm 3.56$ m from the nominal collision vertex in the pseudorapidity intervals $2.09 < \vert \eta \vert < 2.82$ and $2.82 < \vert \eta \vert < 3.84$. CMS has Beam Scintillator Counters (BSC) at  $z = \pm 10.86$  m within $3.23 < \vert \eta \vert < 4.65$. Both experiments also use signals from a beam pick-up based timing system (BPTX) at $z = \pm 175$  m with a time resolution of 200~ps  in their minimum bias trigger.

\begin{figure}[hbt] 
\centering 
\includegraphics[width=0.26\textwidth,keepaspectratio]{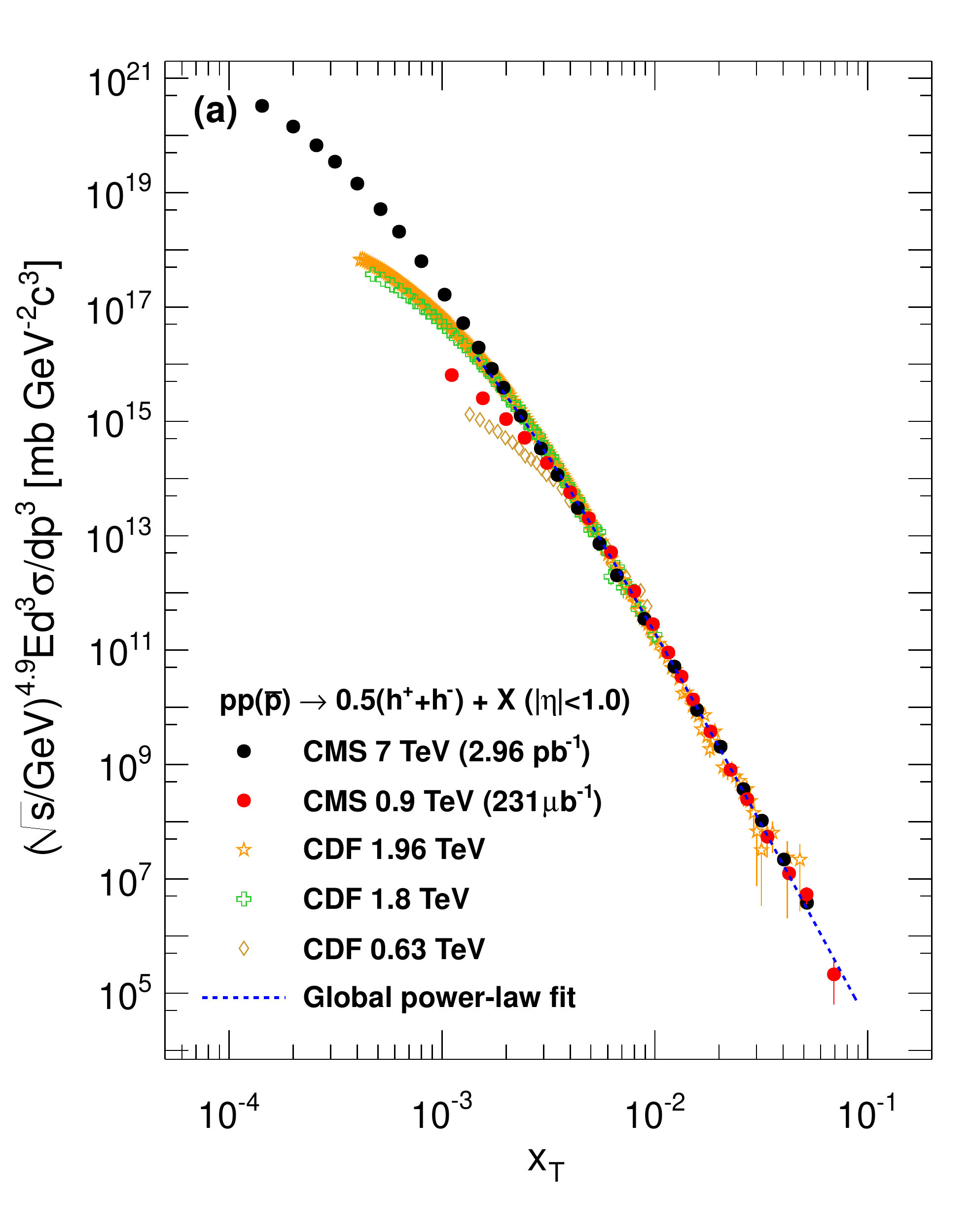}
\hspace{1.5cm}
\includegraphics[width=0.26\textwidth,keepaspectratio]{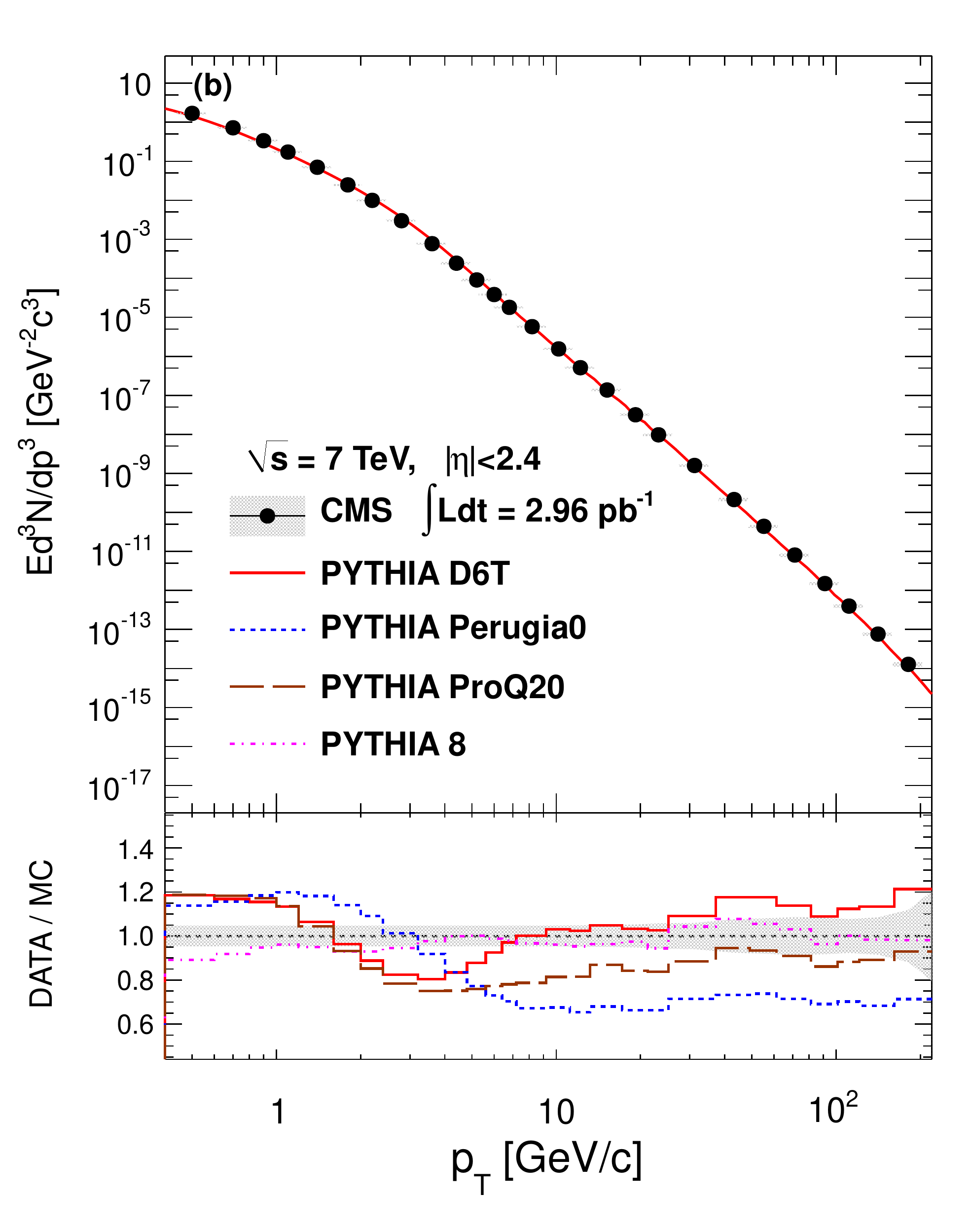}
\vspace{-0.5cm}
\caption{$x_T$ scaling curves (a), inclusive invariant cross-section (b)}
\label{fig:pTspectra}
\end{figure}

Transverse momentum spectra of charged particles have been measured in a large range of $p_T$. Fig.~\ref{fig:pTspectra} shows results from the CMS experiment and comparisons to CDF data \cite{QCD10-008}. Calorimeter-based transverse energy triggers have been used in the high-$p_T$-region instead of the normal minimum bias trigger.
The data are fully corrected.
The inclusive invariant cross-section expressed as a function of the scaling variable $x_T = 2 p_T / \sqrt s$ is given by Eq. \ref{eq:invcross}. 
\begin{equation}
E \frac {d^3\sigma}{dp^3} = F(x_T)/p_T^{n(x_T,\sqrt s)} = F'(x_T)/\sqrt s^{n(x_T,\sqrt s)}
\label{eq:invcross}
\end{equation}

\begin{figure}[hbt] 
\centering 
\includegraphics[width=0.26\textwidth,keepaspectratio]{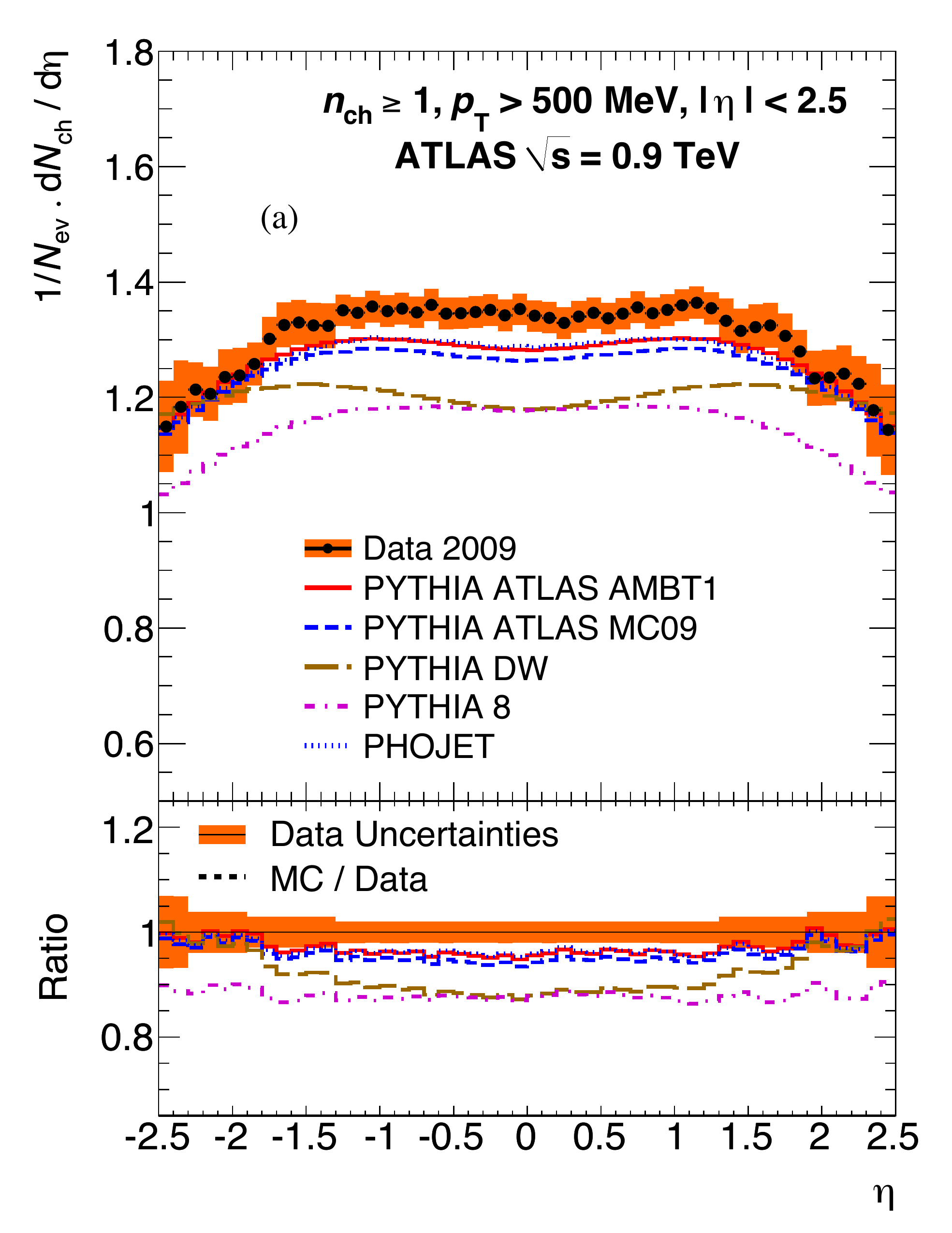}
\includegraphics[width=0.26\textwidth,keepaspectratio]{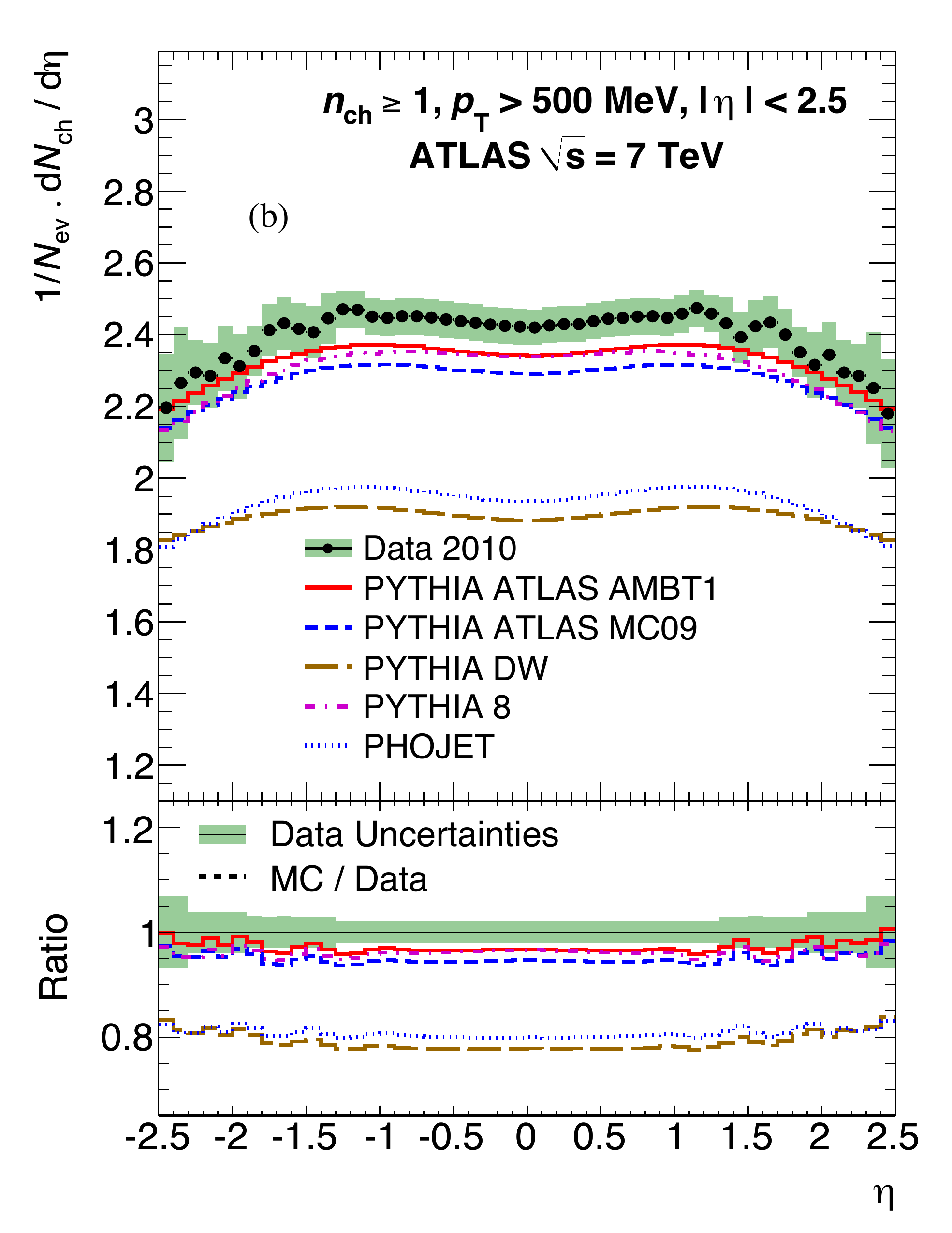}
\includegraphics[width=0.26\textwidth,keepaspectratio]{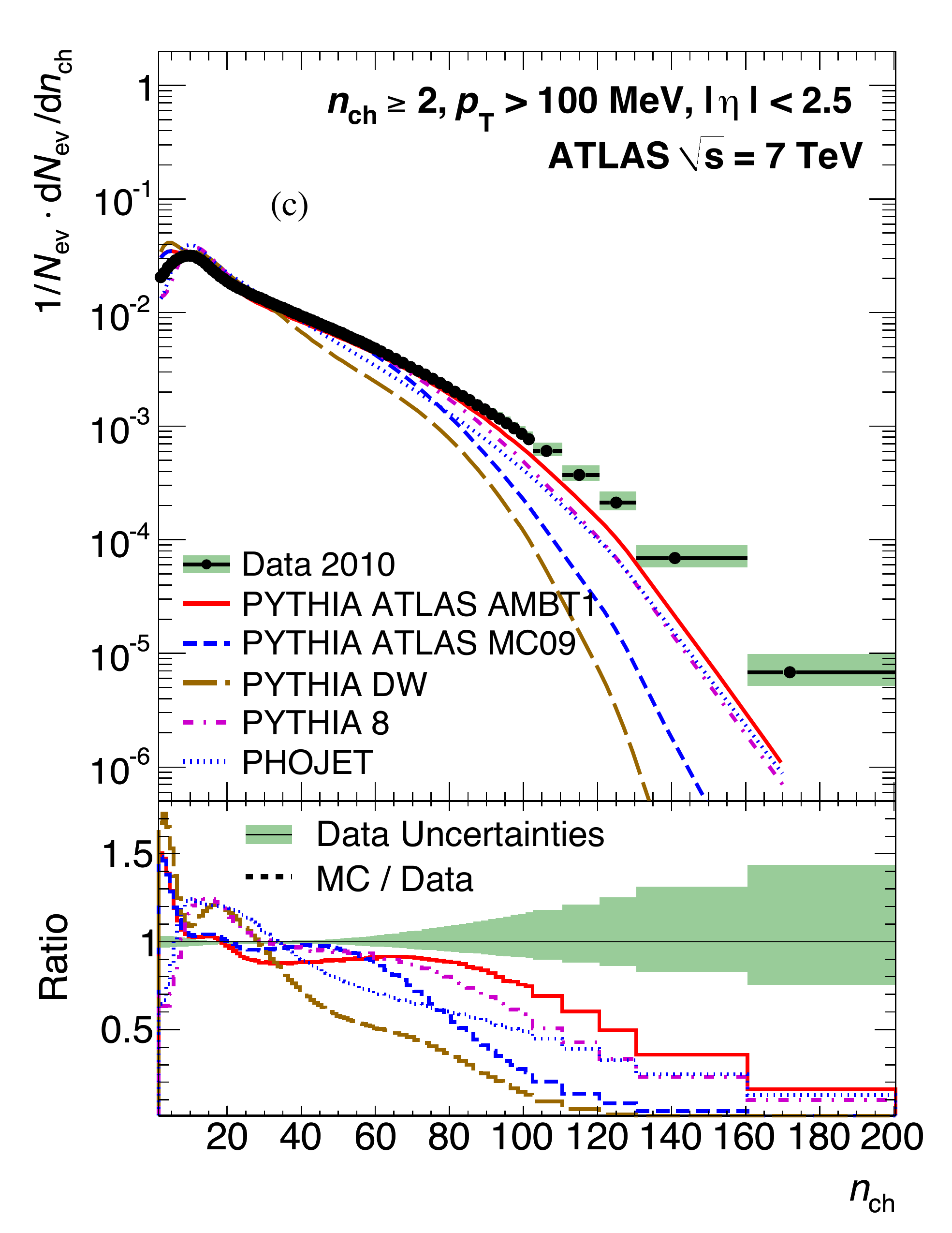}
\caption{Minimum bias pseudorapidity distributions at $\sqrt s$ = 0.9 TeV (a) and 7 TeV (b), multiplicity distribution at $\sqrt s$ = 7 TeV (c)}
\label{fig:rapmultATLAS}
\end{figure}

Minimum bias pseudorapidity and multiplicity distributions as measured by ATLAS \cite{ATLASminbias} are depicted in Fig.~\ref{fig:rapmultATLAS}. The rapidity plateau extends to $\vert \eta \vert \approx 1$ for both centre-of-mass energies of 0.9 and 7 TeV, however, there is an increase of almost a factor of two in its height. No Monte Carlo tune describes the multiplicity distribution shown in Fig.~\ref{fig:rapmultATLAS}c well.

\section{Underlying event studies}
The underlying event (UE) comprises all particles except those from a given hard interaction of interest. It has components from multiple semi-hard parton scattering processes and soft components from beam-beam remnants. The dominant momentum flow defines three regions in the plane transverse to the incoming beams. It is given by the direction of the highest-p$_T$ track in ATLAS, whereas in CMS the leading track jet is used instead. 
The region within an azimuthal angle difference of $ \vert \Delta \phi \vert < 60^0$ with respect to the leading object is called the toward region, and the one opposite ($ \vert \Delta \phi \vert > 120^0$) the away region. The area in between, the transverse region, is the one that is most sensitive to the UE. 

ATLAS \cite{ATLASue} and CMS \cite{CMSue} measured various properties of charged particles in the UE such as multiplicity and transverse momentum distributions. Multiplicity and $\Sigma p_T$ densities as a function of the leading-$p_T$ entity were also studied. 
\begin{figure}[hbt] 
\centering 
\includegraphics[width=0.26\textwidth,keepaspectratio]{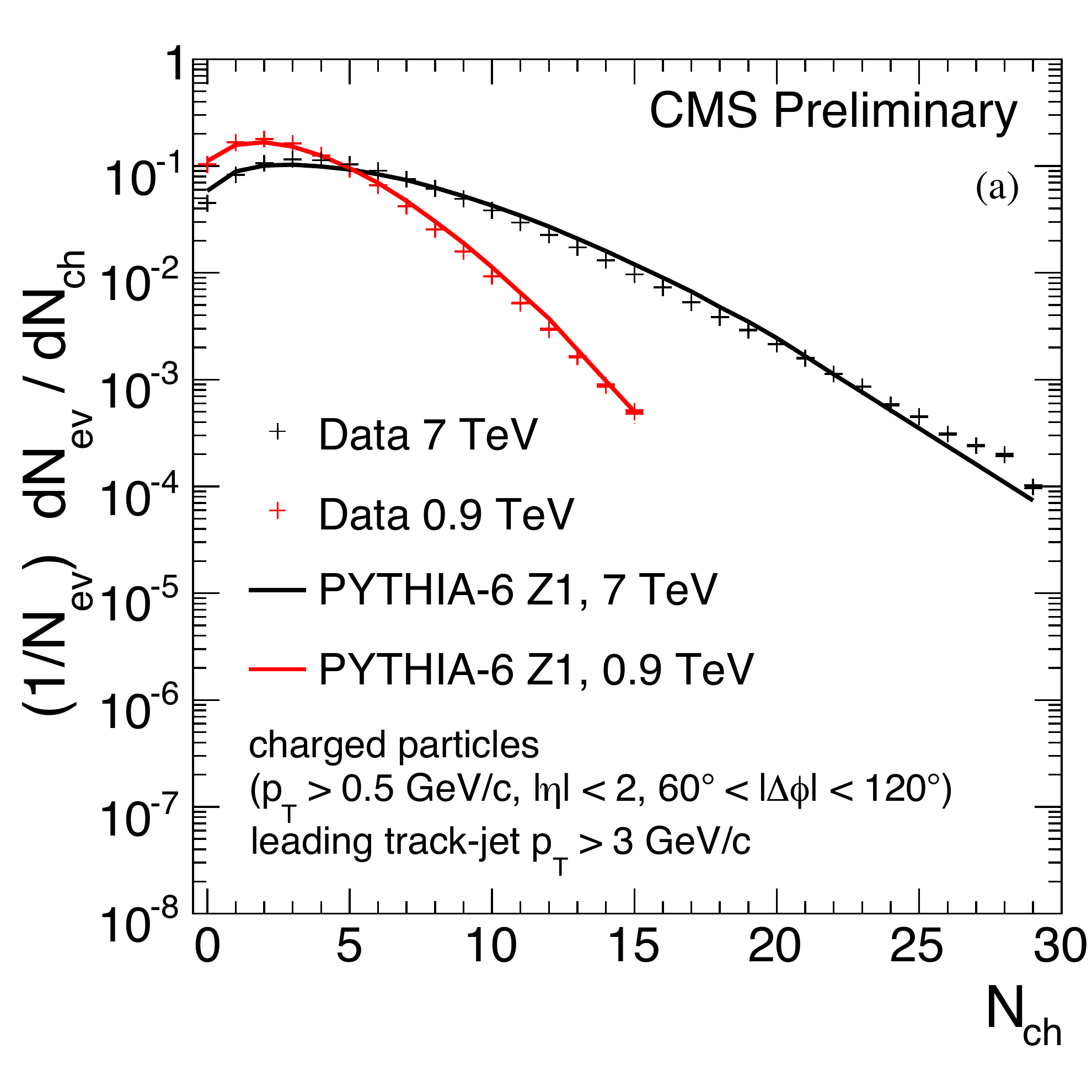}
\hspace{0.7cm}
\includegraphics[width=0.265\textwidth,keepaspectratio]{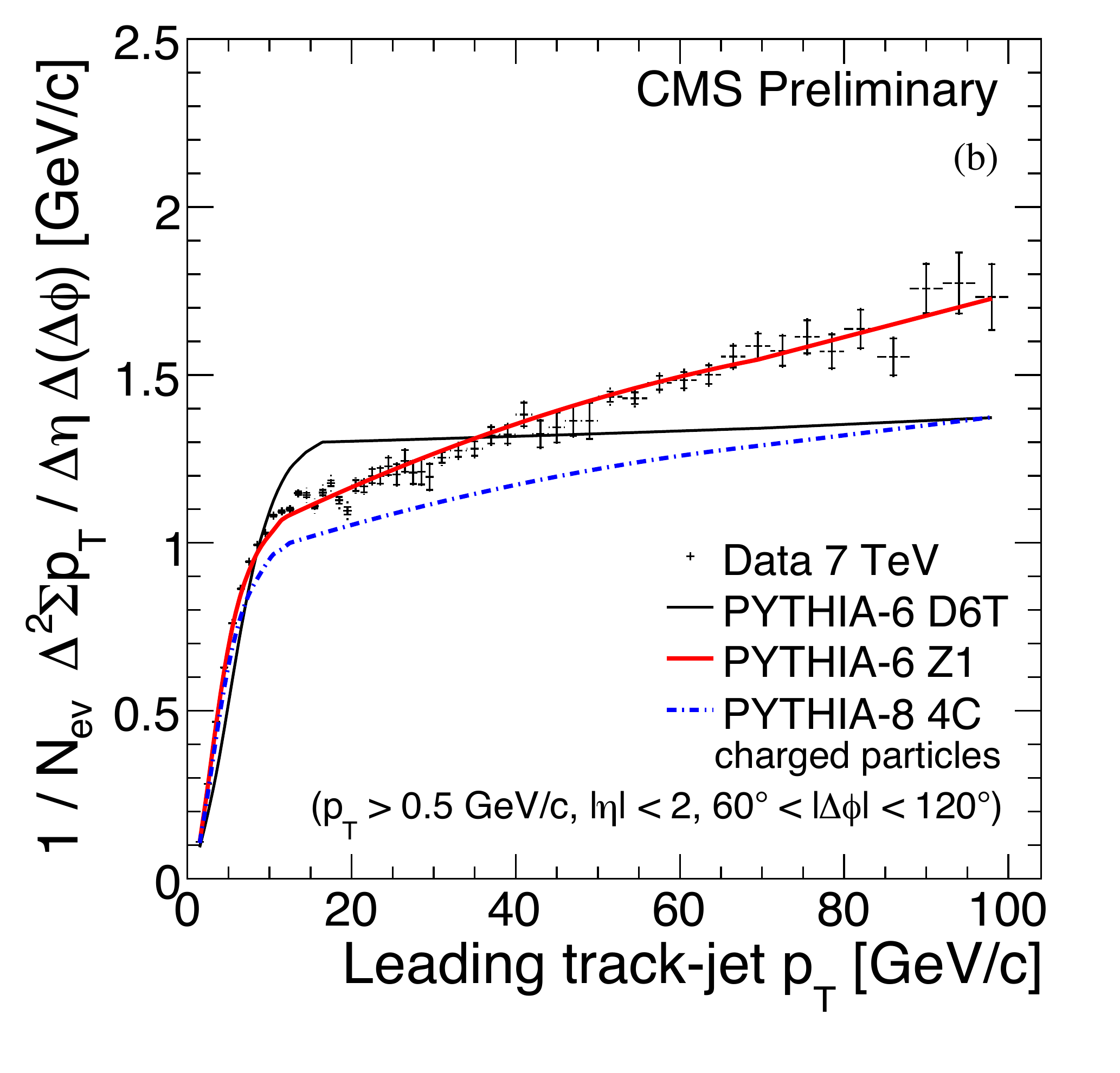}
\hspace{0.7cm}
\includegraphics[width=0.27\textwidth,keepaspectratio]{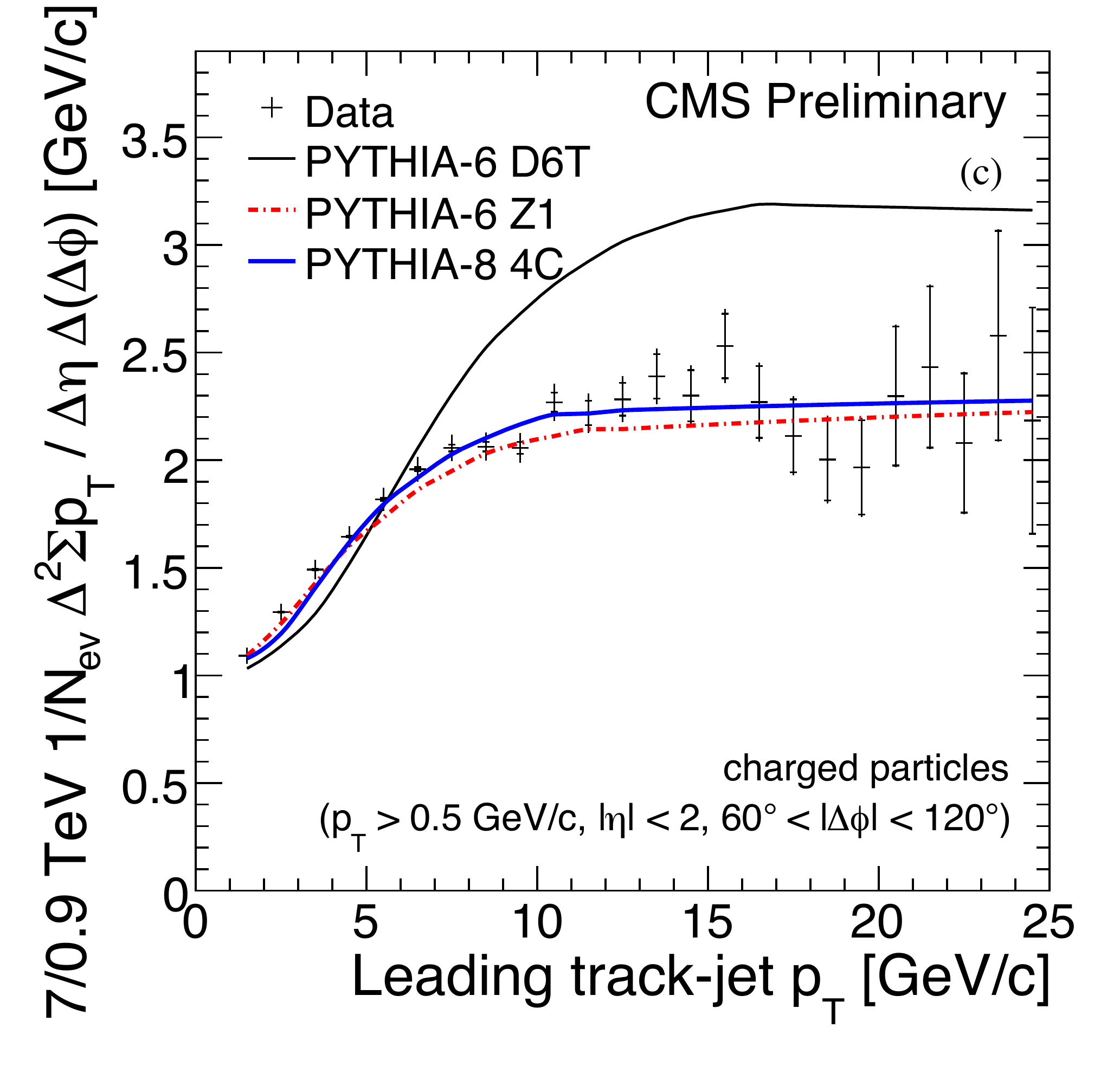}
\caption{Underlying event multiplicites (a), average scalar momentum sum (b), multiplicity density ratio (c)}
\label{fig:NchCMS}
\end{figure}
There is a strong growth of UE activity with $\sqrt s$. Fig.~\ref{fig:NchCMS}a shows the increase of the number of events for centre-of-mass energies from  0.9 to 7 TeV as a function of multiplicity for the transverse region. The average scalar momentum sum rises sharply till 8 GeV due to the increase of multiple parton interaction activity, followed by a slow increase thereafter (Fig.~\ref{fig:NchCMS}b). 
The distributions in Fig.~\ref{fig:NchCMS} are 
well reproduced by the PYTHIA Z1 Monte Carlo tune \cite{PYTHIAZ1}.

\begin{figure}[hbt] 
\centering 
\includegraphics[width=0.3\textwidth,keepaspectratio]{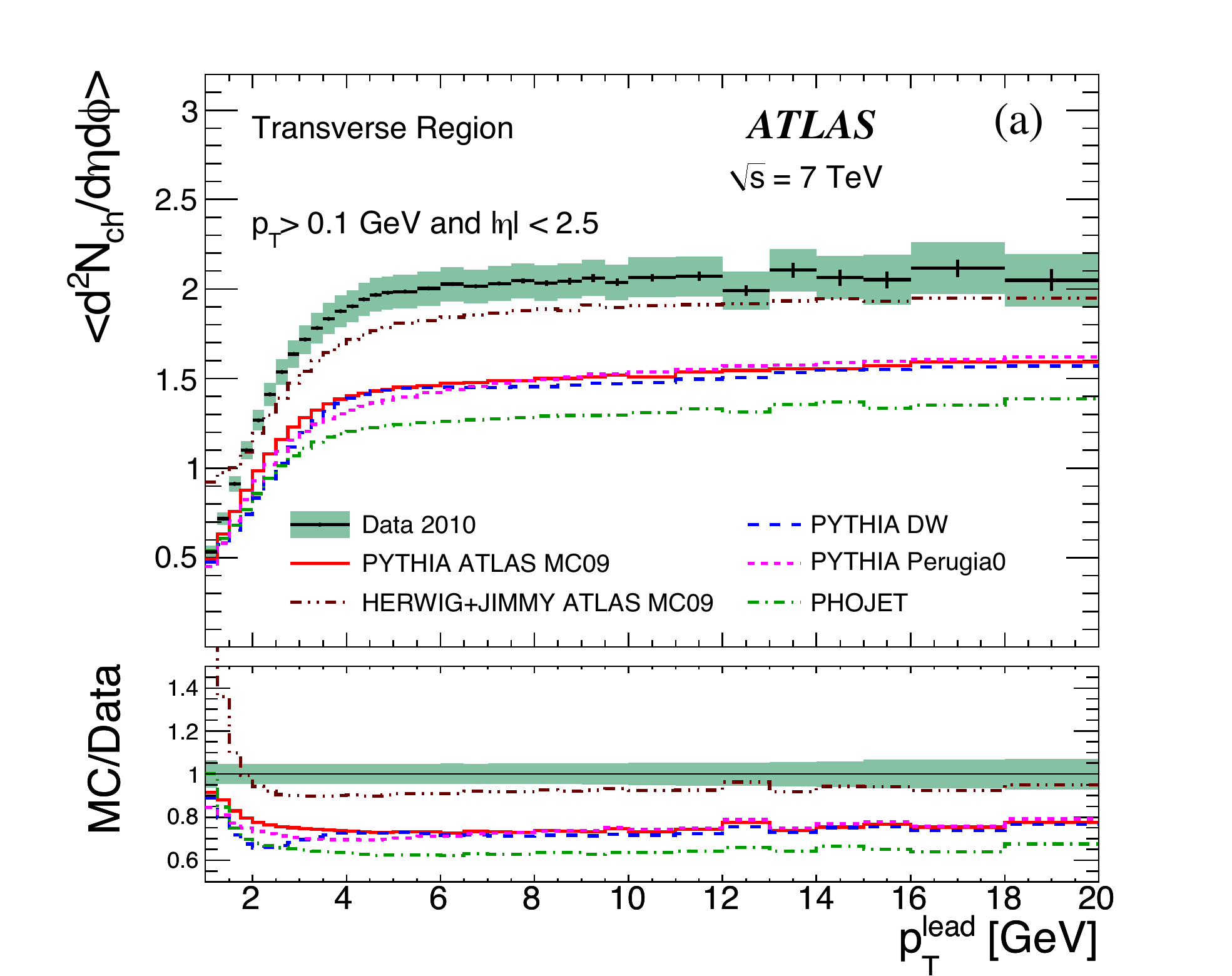}
\includegraphics[width=0.33\textwidth,keepaspectratio]{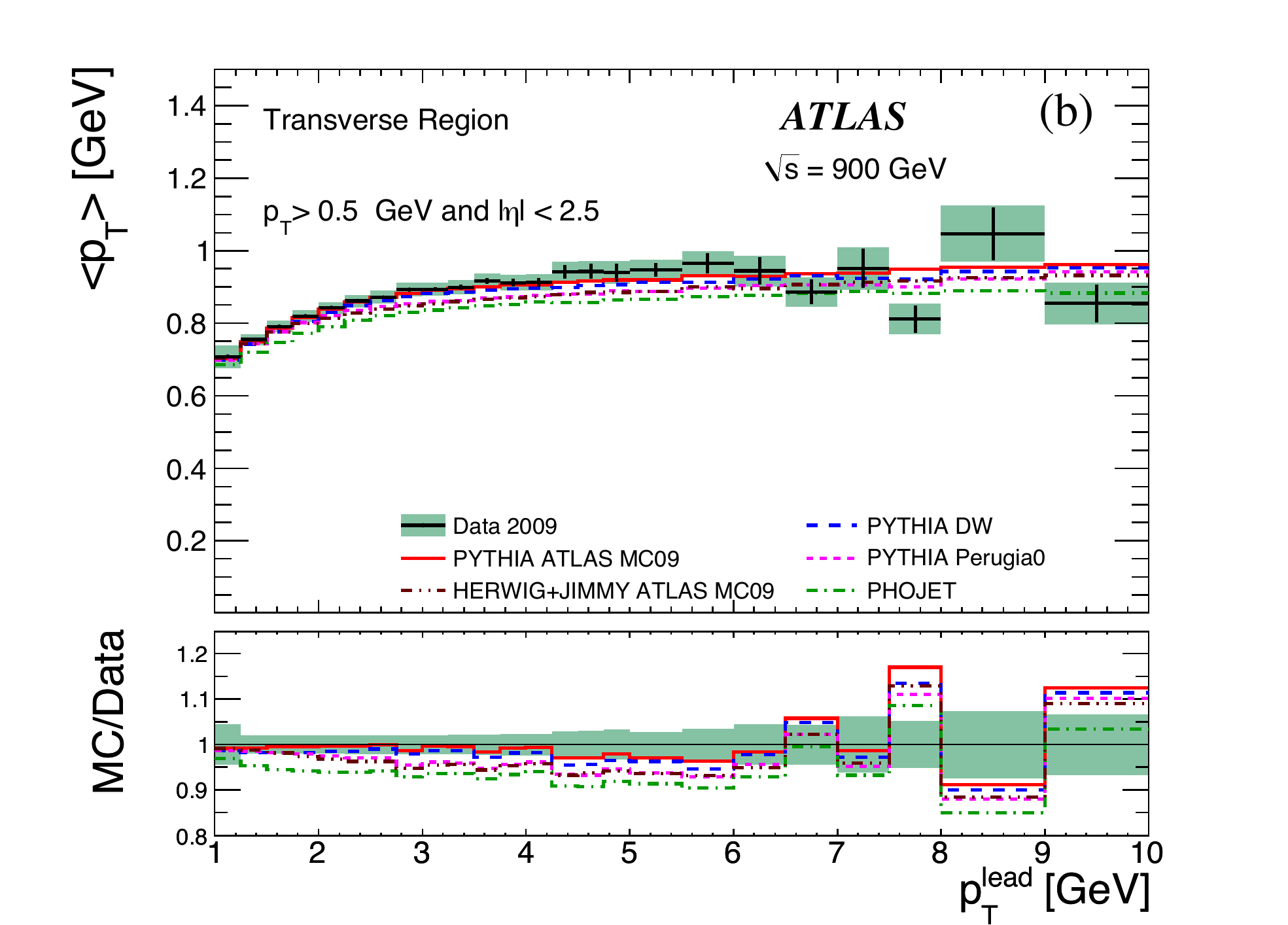}
\includegraphics[width=0.33\textwidth,keepaspectratio]{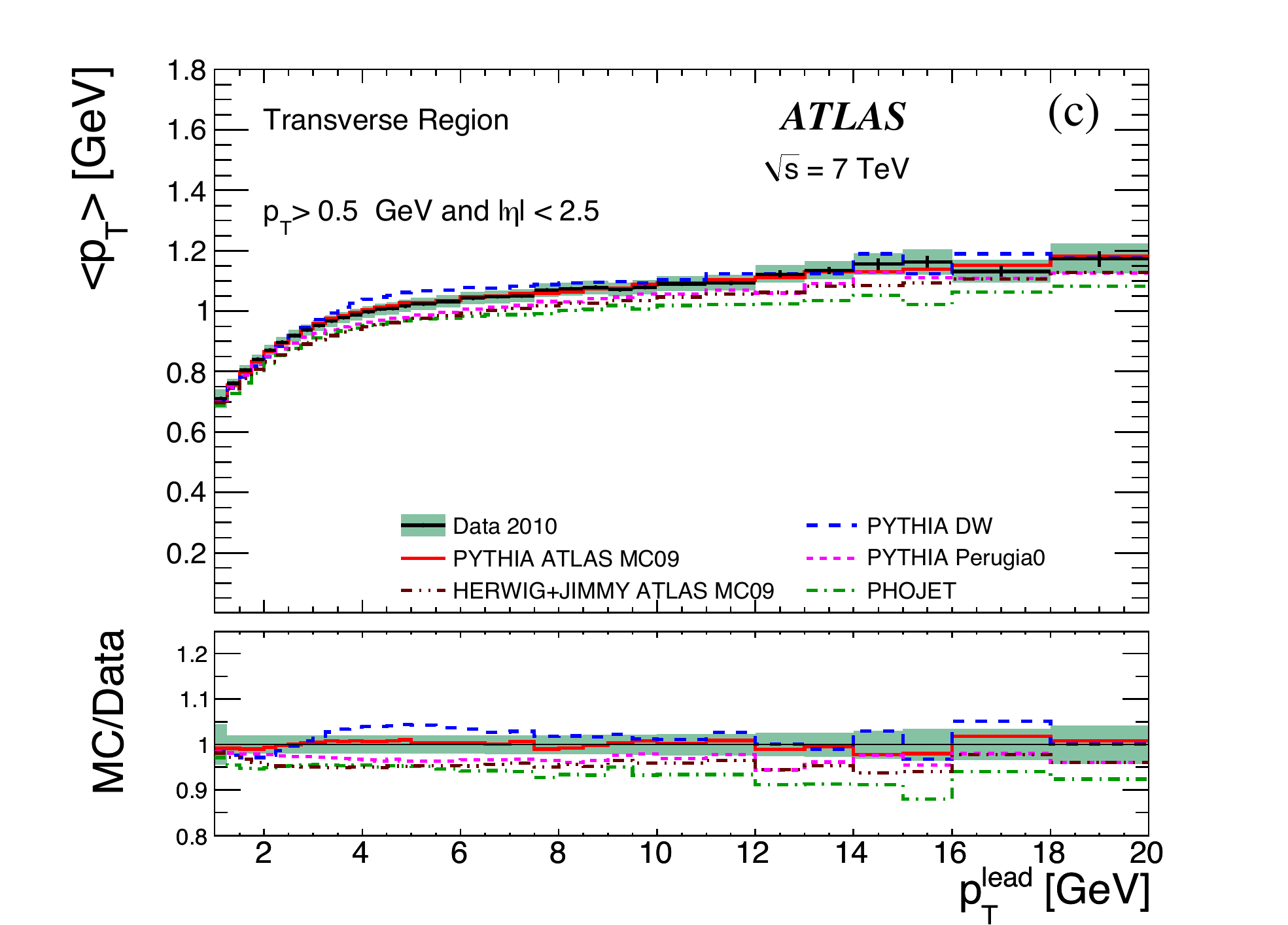}
\caption{Charged particle multiplicity density (a), average transverse momentum at $\sqrt s$ = 900~GeV (b) and 7~TeV (c)}
\label{fig:ATLASdistributions}
\end{figure}

The charged particle multiplicity density in the transverse region is plotted in Fig.~\ref{fig:ATLASdistributions}a. Compared to minimum bias there is about two times more activity in the UE. All Monte Carlo models underestimate the multiplicity. Figs.~\ref{fig:ATLASdistributions}b and  \ref{fig:ATLASdistributions}c show the increase of the UE $p_T$ by about 20\% from 0.9 to 7 TeV, well reproduced by a variety of models.

\section{Strangeness production}
\begin{figure}[hbt] 
\centering 
\includegraphics[width=0.325\textwidth,keepaspectratio]{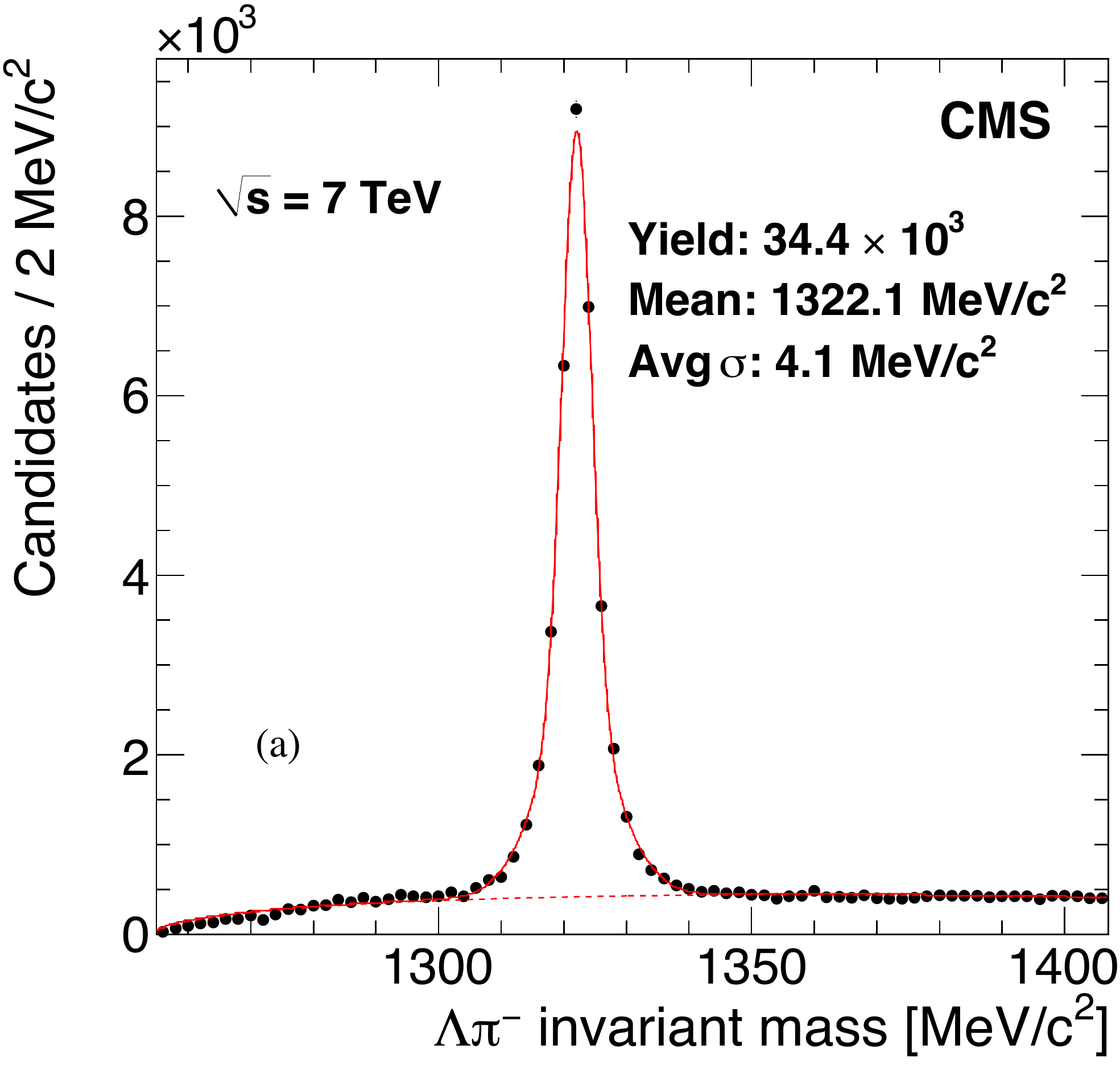}
\includegraphics[width=0.31\textwidth,keepaspectratio]{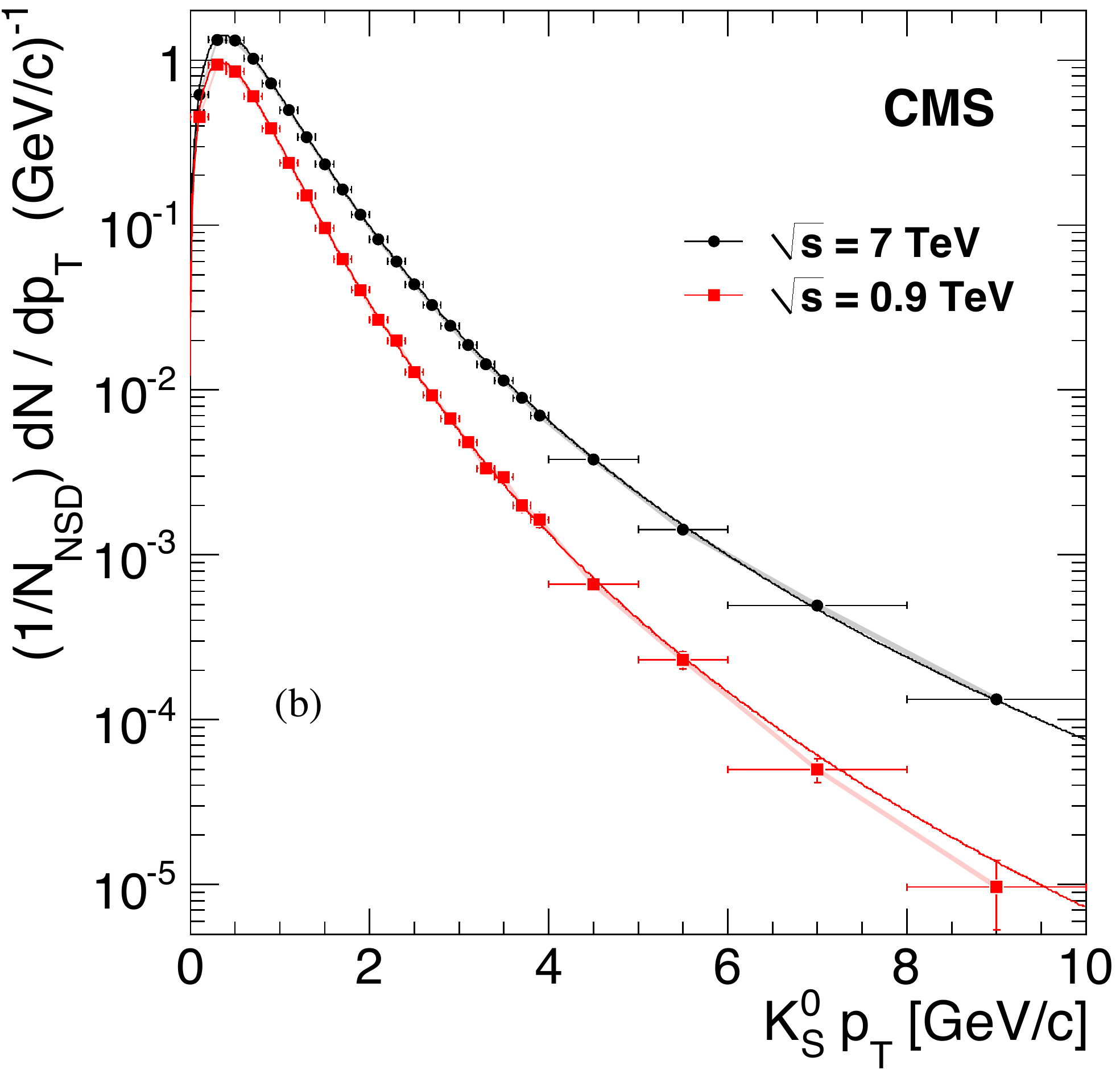}
\includegraphics[width=0.24\textwidth,keepaspectratio]{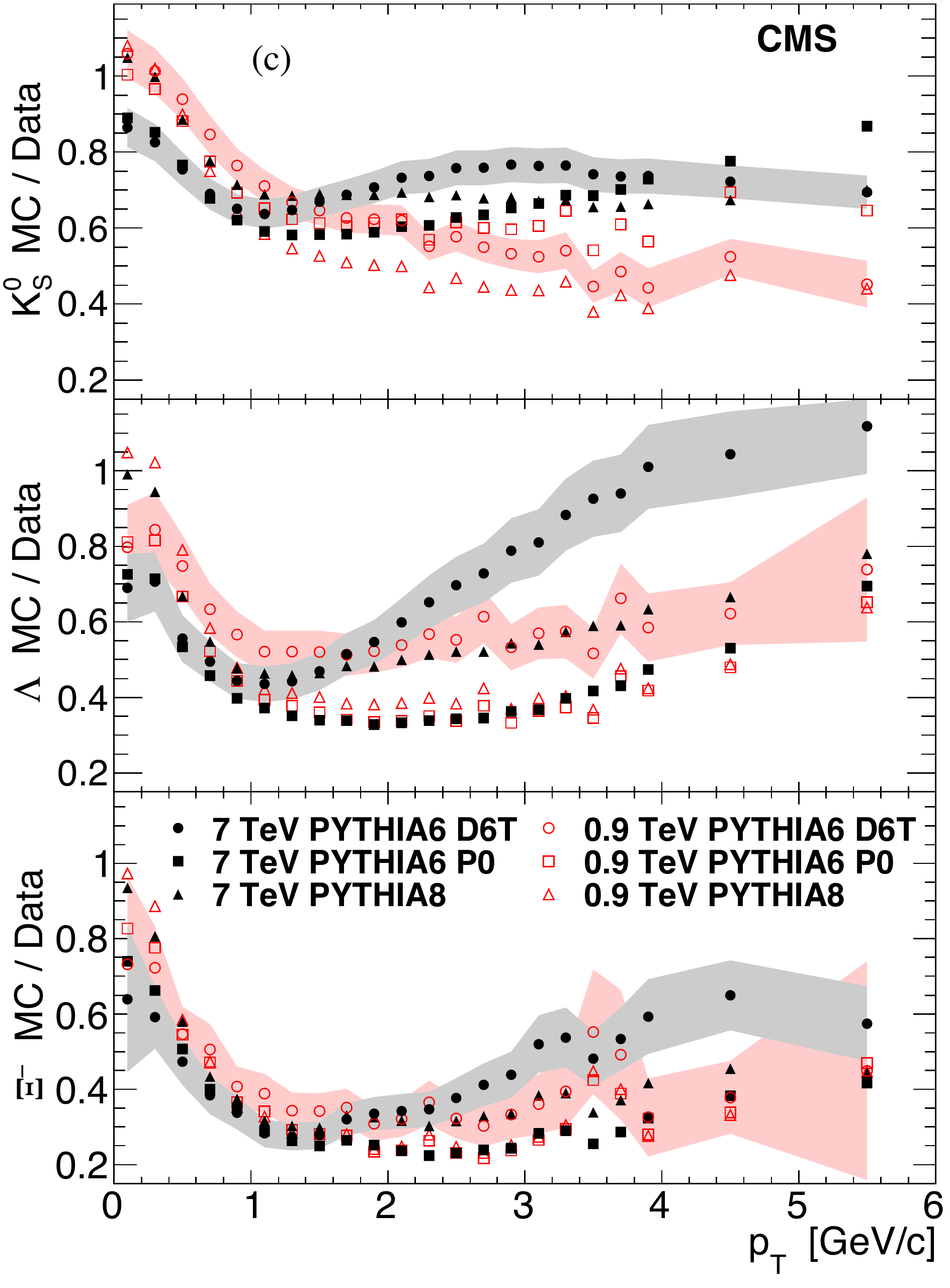}
\caption{$\Xi^-$ mass peak (a),  $p_T$ distribution of $\PKzS$ (b), hyperon $p_T$ yield ratios(c)}
\label{fig:strange}
\end{figure}

The production of the strange hadrons $\PKzS, \Lambda$ and $\Xi$ has been studied \cite{strange}. The mass peaks have been reconstructed, as shown in Fig.~\ref{fig:strange}a for the $\Xi^-$. 
Production rates have been measured as functions of rapidity and transverse momentum. The $p_T$ distributions extend from practically zero to 10 GeV for $\PKzS$ (Fig.~\ref{fig:strange}b) and 6 GeV for $\Xi^-$. 
The increase in production of strange particles from 0.9 to 7 TeV is approximately consistent with results for charged particles described above, but the rates exceed the predictions by up to a factor of three (Fig.~\ref{fig:strange}c). This deficiency probably originates from parameters regulating the frequency of s-quarks appearing in color strings. 

\section{Particle correlations}

\begin{figure}[hbt] 
\centering 
\includegraphics[width=0.65\textwidth,keepaspectratio]{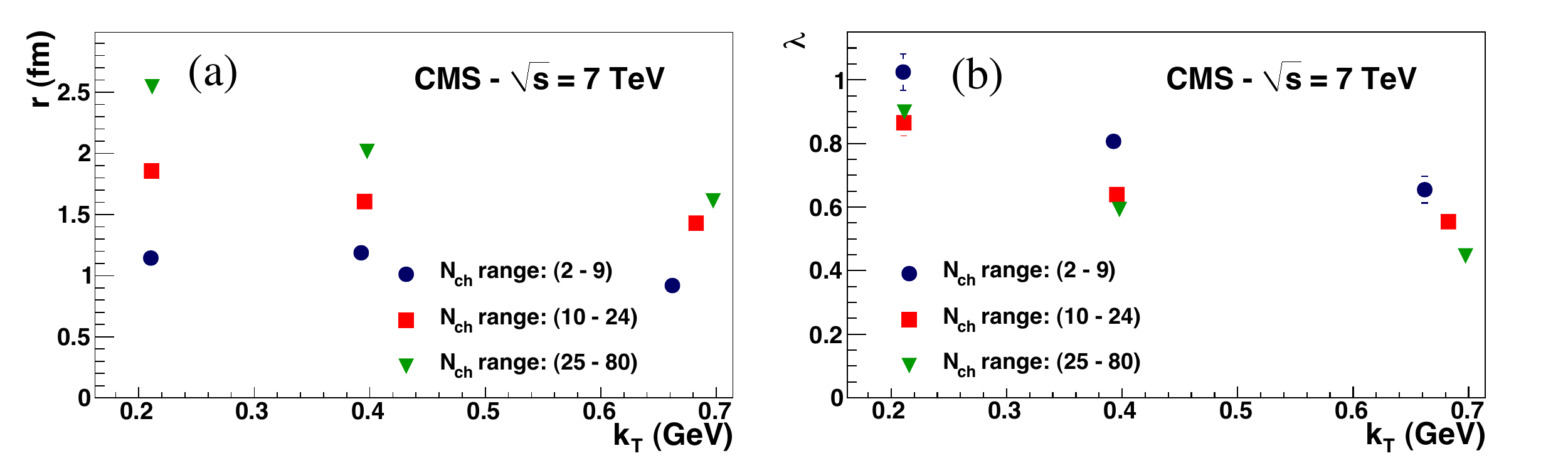}
\includegraphics[width=0.3\textwidth,keepaspectratio]{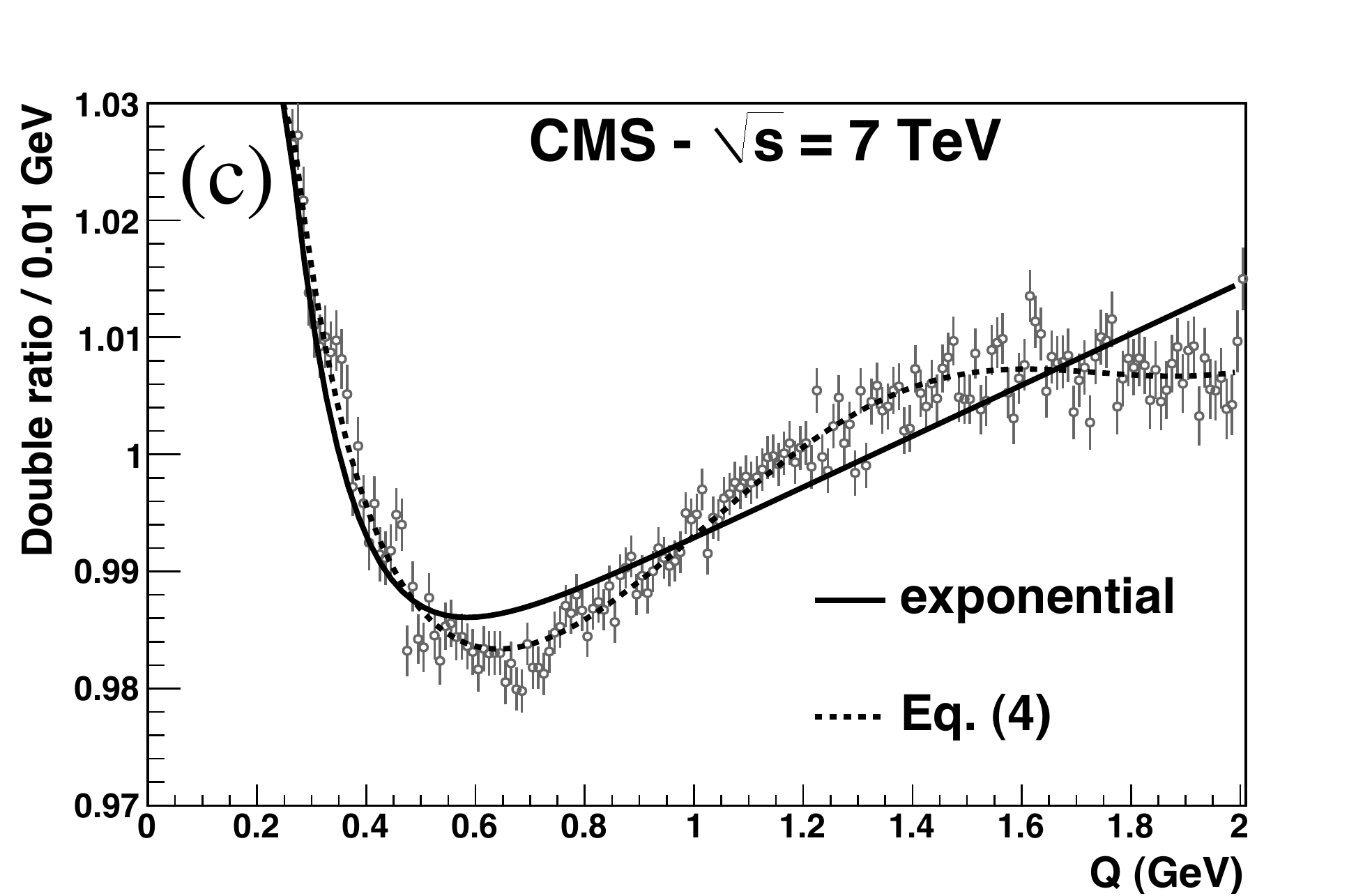}
\vspace{-0.2cm}
\caption{Radius r (a), correlation strength $\lambda$ (b),  double ratio showing anticorrelations (c)}
\label{fig:BE}
\end{figure}

Pairs of same-sign charged particles with four-momentum difference $Q$ in the region 0.02 GeV~$< Q < $ 2 GeV are analysed to study Bose-Einstein correlations \cite{BE}.
The radius of the effective space-time region emitting bosons with overlapping wave functions increases with multiplicity, whereas the correlation strength decreases. Both decrease with increasing momentum difference $k_T$.
Anticorrelations between same-sign charged particles are observed for $Q$ values above the signal region, which can be seen in 
Fig.~\ref{fig:BE}c showing the double ratio $R(Q)$ \cite{TauModel}.

\begin{figure}[hbt] 
\centering 
\includegraphics[width=0.65\textwidth,keepaspectratio]{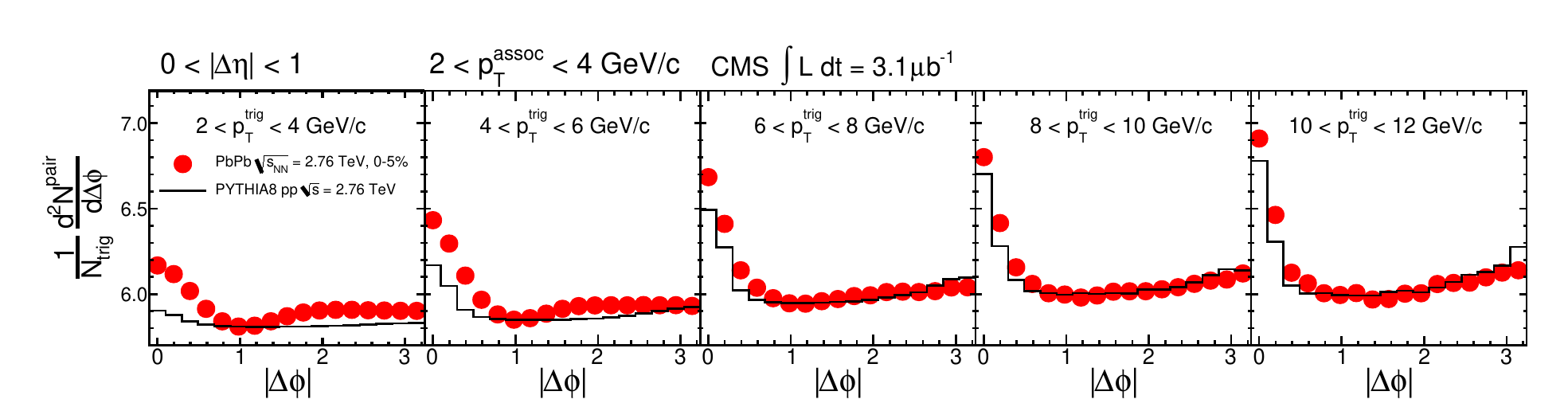}
\includegraphics[width=0.65\textwidth,keepaspectratio]{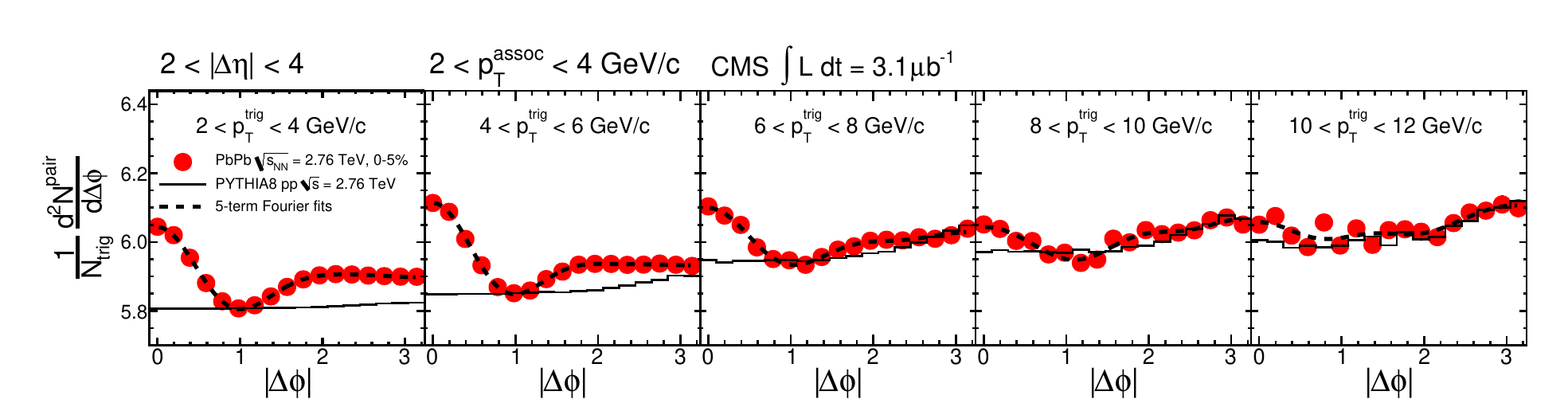}
\vspace{-0.3cm}
\caption{Short-range ($0 < \vert \Delta \eta \vert < 1$) and long-range ($2 < \vert \Delta \eta \vert < 4$) correlations in lead ion data}
\label{fig:ionridge}
\end{figure}

Near-side long-range particle correlations in proton data have first been reported by CMS~\cite{Ridge}. A ridge, a pronounced structure in high-multiplicity events for rapidity and azimuth differences of $2.0 < \vert \Delta \eta \vert < 4.8$ and $\Delta \phi  \approx 0$ has been found. Long- and short-range correlations in ion data have also been studied \cite{AziCorrPbPb}. 
The correlation functions of the 0-5\% most central collisions show characteristic features not present in minimum bias proton interactions (Fig.~\ref{fig:ionridge}). The ridge is most evident for $p_T$'s of the trigger particle between 2 and 6 GeV, but disappears at high $p_T$.

\end{document}